\documentclass[float,epsfig,usenatbib]{mn2e}

\usepackage{graphicx}
\usepackage{appendix}
\usepackage{psfrag}
\usepackage{amsmath,amssymb}
\usepackage{threeparttable}
\usepackage{float}
\usepackage{subfigure}
\usepackage{afterpage}

\title[The migration of nearby spirals]{The migration of nearby spirals from the blue
to red sequence: AGN feedback or environmental effects?}
\author[T.M. Hughes \& L. Cortese]
{T.M. Hughes $^{1}$\thanks{thomas.hughes@astro.cf.ac.uk} and L. Cortese$^{1}$.
\\
$^{1}$School of Physics and Astronomy, Cardiff University, Queens Buildings, The Parade, Cardiff, CF24 3AA, UK\\}

\date{Accepted for publication in MNRAS}

\begin{document}
\newcommand{\Zsolar}{\mbox{$\,\rm Z_{\odot}$}}
\newcommand{\Msolar}{\mbox{$\,\rm M_{\odot}$}}
\newcommand{\Lsolar}{\mbox{$\,\rm L_{\odot}$}}
\newcommand{\xs}{$\chi^{2}$}
\newcommand{\dxs}{$\Delta\chi^{2}$}
\newcommand{\xsn}{$\chi^{2}_{\nu}$}
\newcommand{\ls}{{\tiny \( \stackrel{<}{\sim}\)}}
\newcommand{\gs}{{\tiny \( \stackrel{>}{\sim}\)}}
\newcommand{\asec}{$^{\prime\prime}$}
\newcommand{\amin}{$^{\prime}$}
\newcommand{\mstar}{\mbox{$M_{*}$}}
\newcommand{\hi}{H{\sc i}\ }
\newcommand{\hii}{H{\sc ii}\ }
\maketitle

\label{firstpage}

\begin{abstract}

We combine ultraviolet to near-infrared photometry with \hi 21cm line observations for a complete volume-limited sample of nearby galaxies in different environments (from isolated galaxies to Virgo cluster members), to study the migration of spirals from the blue to the red sequence.
Although our analysis confirms that, in the transition region between the two sequences, a high fraction of spirals host active 
galactic nuclei (AGN), it clearly shows that late-types with quenched star formation are mainly \hi deficient galaxies preferentially found in the Virgo cluster. 
This not only suggests that environmental effects could play a significant role in driving the migration of local galaxies from the blue sequence, 
but it also implies that a physical link between AGN feedback and quenching may not be assumed from a correlation between nuclear activity and colour.

\end{abstract}


\section{Introduction}

It is well known that galaxies form two sequences in the colour-magnitude space: star-forming disks occupy the blue cloud whereas quiescent, bulge-dominated objects reside on the red sequence (e.g. \citealp{tully1982}). 
The origin of this bimodality in the colour distribution of galaxies is still a puzzle.
There is mounting evidence that, since z$\sim$1, the stellar mass density of the red sequence has increased by at least a factor $\sim$2, 
while surprisingly the stellar mass density of the blue cloud has remained constant, despite continued star formation 
over several Gyr \citep{bell2007,faber2007}. 
A possible explanation of this is that galaxies form most of their stars while they are in the blue cloud but then migrate to the red sequence, gradually increasing the stellar mass retained in quiescent systems.
In this case, a quenching of the star formation is required to drive galaxies towards the red sequence, but 
it is still unclear what physical mechanism(s) may be responsible for such migration.

The most popular candidate is feedback from the accretion of material onto supermassive black holes (i.e., AGN feedback). 
The AGN may heat or expel the surrounding gas, thus preventing star formation (e.g., \citealp{croton2006}). 
Theoretical studies have shown that models including AGN feedback provide a better match between theory and observations (e.g. the bright end of the luminosity function), making this quenching mechanism very promising \citep{baugh2006}. 
Recently, this hypothesis has gained additional support from observations showing that the AGN-host fraction peaks in the region between the blue and red sequence (i.e., the transition region or {\it green valley}, \citealp{martin2007,schawinski2007}).
However, it is still a matter of debate whether these results imply a physical connection between AGN activity and suppression of the star formation in galaxies (e.g. \citealp{geo2008}).
In addition, though AGN feedback has been directly observed in the giant ellipticals in the centre of clusters (e.g. \citealp{forman2007}), it is unclear whether this mechanism works in spirals \citep{okamoto2008}, which typify the transition galaxies today. 

In contrast to the above internal mechanism, environmental effects may be responsible for quenching the star formation activity, driving the migration  of spiral galaxies out of the blue cloud.
Gravitational interactions \citep{merritt1984,moore1996}, ram pressure stripping \citep{gunngott1972}, and hybrid processes combining multiple mechanisms such as preprocessing \citep{fujita2004,cortese2006b}, can in fact quench star formation in galaxies in high density environments (see \citealp{boselli2006} for a detailed review of these processes). This is clearly reflected in the morphology-density \citep{dressler1980,whitmore1993}, star formation-density (e.g. \citealp{kennicutt1983}) and gas-density relations \citep{haynes1984} observed in the local universe. In addition, it has recently been shown that the environment is responsible for the formation of the red sequence at low luminosities \citep{boselli2008}. 
Discriminating between the effects of AGN feedback and environment on star formation is therefore 
necessary to unravel the evolutionary history of transition galaxies.


In this Letter, we attempt to assess whether the mechanism suppressing the star formation in nearby spirals is an internal or environmental process. We use a complete volume-limited sample of galaxies covering different environments, from galaxies in isolation to the core of the Virgo cluster.
For the first time, we combine near-infrared and optical imaging and optical spectroscopy with \hi 21cm line and ultraviolet data, which trace the galaxy gas content and current star formation, respectively.

\section{The Sample}

\begin{table}
\begin{center}
\begin{tabular}{l c c c c}
\cline{1-5}
\\
Sample & NUV & Spectroscopy & \hi \\
\cline{1-5}
\\
Total & 86.9\%  &  76.4\%  & 86.6\%  \\ 
   &(394/454)&(347/454)&(393/454) \\
Early-type & 100.0\%  &  - & -  \\ 
                   & (171/171) &  &  \\
Late-type & 91.2\%  & 86.6\%  & 96.8\%  \\ 
                  &(258/283)&(245/283)&(274/283) \\
 & & & & \\
Late-type & 94.5\%  &  97.9\%  & 100.0\%  \\ 
(Cluster)       &(136/144)&(143/144)&(144/144) \\
Late-type & 87.8\%  &  73.4\%  & 93. 5\%  \\ 
(Field)       &(122/139)&( 102/139)&(130/139) \\
\\
\cline{1-5}
\end{tabular}
\caption{The completeness of the NUV data, nuclear spectroscopy and \hi observations for the total sample and for the different subsamples used in this paper.}
\label{tab:compstats}
\end{center}
\end{table}

We selected a complete volume-limited sample following the criteria of the Herschel Reference Survey (HRS; \citealp{boselli2009}), a guaranteed-time key project with the Herschel Space Telescope.
In detail, we select objects with 2MASS \citep{jarrett2003} K-band magnitude K$_{Stot} \le$ 12 mag and with optical recessional velocity between 1050 km s$^{-1}$ and 1750 km s$^{-1}$, corresponding to a distance range of 15-25 Mpc, assuming a Hubble constant H$_{0} =\ $70 km s$^{-1}$ Mpc$^{-1}$ and no peculiar motions. In the Virgo Cluster, where peculiar motions are dominant, we use the subgroup membership and distances as determined in \cite{gav1999}. Additionally, we select galaxies at high galactic latitudes (b $>$ +55$^{\circ}$) and inhabiting regions of low galactic extinction, A$_{B}$ $<$ 0.2 \citep{schlegel1998}, to minimize galactic cirrus contamination.
The total sample contains 454 galaxies of which 171 are early-types\footnote{We note that we select more early-type galaxies compared to the HRS, which only includes early-type galaxies having K$_{Stot} \le$ 8.7 mag.} (from dE to S0/Sa) and 283 are late-type (Sa and later types) systems\footnote{We exclude VCC1327 from further analysis due to the presence of a bright star superimposed on the galaxy image (as noted in the literature, e.g. \citealp{binggeli1985}) making any accurate photometry difficult.}. 
Morphological classifications are taken either from the Virgo Cluster Catalogue (VCC; \citealp{binggeli1985}), NED or the RC3 \citep{vauc1991}. 
A visual inspection of the SDSS images was used for a few cases where morphological classifications were unavailable. 

We collected multi-wavelength data from a variety of sources. Observations from the GALEX \citep{martin2005} GR2 to GR4 data releases in the near-ultraviolet (NUV; $\lambda$=2316 \AA: $\Delta \lambda$=1069 \AA) band were available for 394 objects. NUV magnitudes were obtained by integrating the flux over the galaxy optical size, determined at the surface brightness of $\mu$(B) = 25 mag arcsec$^{-2}$. The magnitudes are accurate to $\pm 10 \% $. 
Optical B, V and near-infrared H band photometry was taken from the GOLDMine database \citep{gav2003} and 2MASS, respectively. 
All magnitudes have been corrected for Galactic extinction according to \cite{schlegel1998}.
Internal dust attenuation was determined using the total infrared (TIR) to UV luminosity ratio method (e.g., \citealp{xu1995}) and the age-dependent relations 
of \cite{cortese2008}. The TIR luminosity is obtained from IRAS 60 and 100 $\mu$m fluxes or, in the few cases when IRAS observations are not available, using 
the empirical recipes described in \cite{cortese2006a}. The typical uncertainty in the NUV dust attenuation is $\sim$ 0.5 mag. 
The internal dust attenuation at optical and infrared wavelengths were then derived assuming the Large Magellanic Cloud extinction law as described in \cite{cortese2008}.
In order to convert H-band luminosities into stellar mass (\mstar), we adopted the $B-V$ colour-dependent stellar mass-to-light ratio relation from 
\cite{bell2003}, assuming a \cite{kroupa1993} initial mass function. Morphologically averaged $B-V$ colours were used when optical observations were unavailable. 

We combined optical spectroscopy from the Sloan Digital Sky Survey (SDSS; \citealp{sdss2000}) Data Release 7 and \cite{decarli2007}, with nuclear classification 
available from NED to investigate the AGN activity of our sample.    
Data was available for 347 objects, thus giving a completeness of $\sim$76\%. 
Following \cite{decarli2007}, we divided galaxies into three different classes based on the following optical line flux ratios: 
 $[NII] / H\alpha < 0.4$ for those containing \hii star-forming regions, $[NII] / H\alpha > 0.6$ for those displaying AGN-like behavior and 
 $0.4 < [NII] / H\alpha < 0.6$ for those showing both AGN and star forming activity.

Single-dish \hi 21 cm line emission data, necessary for quantifying the gas content of galaxies, was taken from 
\cite{springob2005} and \cite{gav2003}. 
Overall, 393 objects have \hi data. We estimate the \hi deficiency parameter ($HI_{DEF}$) as defined by \cite{haynes1984}, 
using the morphologically-dependent coefficients presented by \cite{solanes1996}. 
In the following, galaxies are classified as \hi deficient if $HI_{DEF}> 0.5$ 
(i.e., they have lost $\geq$70\% of their original atomic hydrogen when compared to 
isolated galaxies of the same size and morphological type).   

The completeness in NUV, spectroscopic and \hi data of our sample is summarized in Table \ref{tab:compstats}.
Late-type galaxies without nuclear classification or \hi observations are considered as non-active, gas-rich systems in the rest of this work.  
 
Two subsamples were created based on the environment inhabited by each object in the sample. 
All galaxies with membership of the Virgo cluster, as defined by \cite{gav1999}, were selected for the `cluster' subsample. 
Those galaxies outside Virgo, which range from isolated systems to galaxies in groups, were assigned to the `field' subsample.
The statistics of the late-type galaxies in the cluster and field samples are also presented in Table~\ref{tab:compstats}.   

\begin{figure*}
\begin{center}
\includegraphics[width=0.32\textwidth]{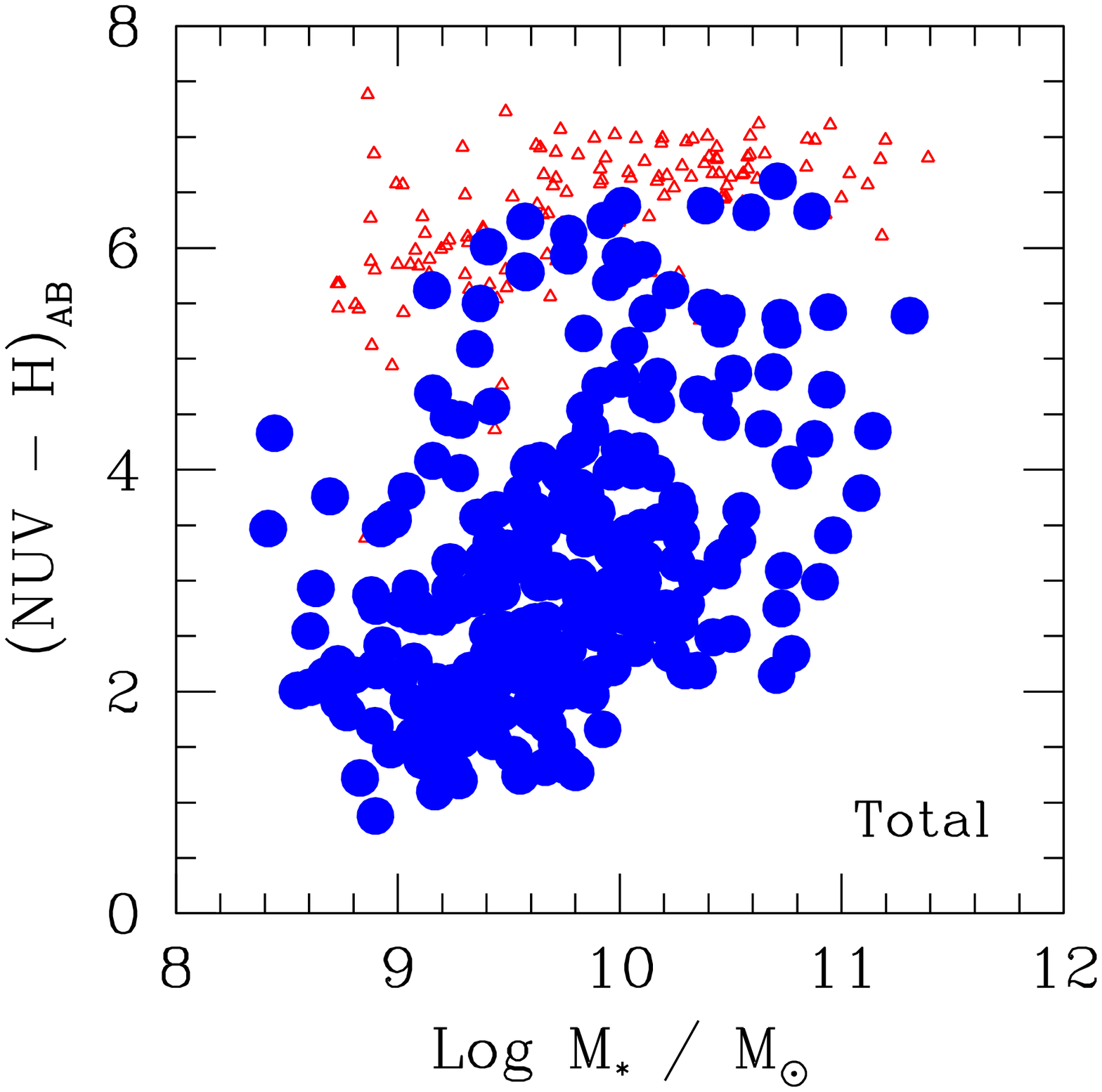}
\includegraphics[width=0.32\textwidth]{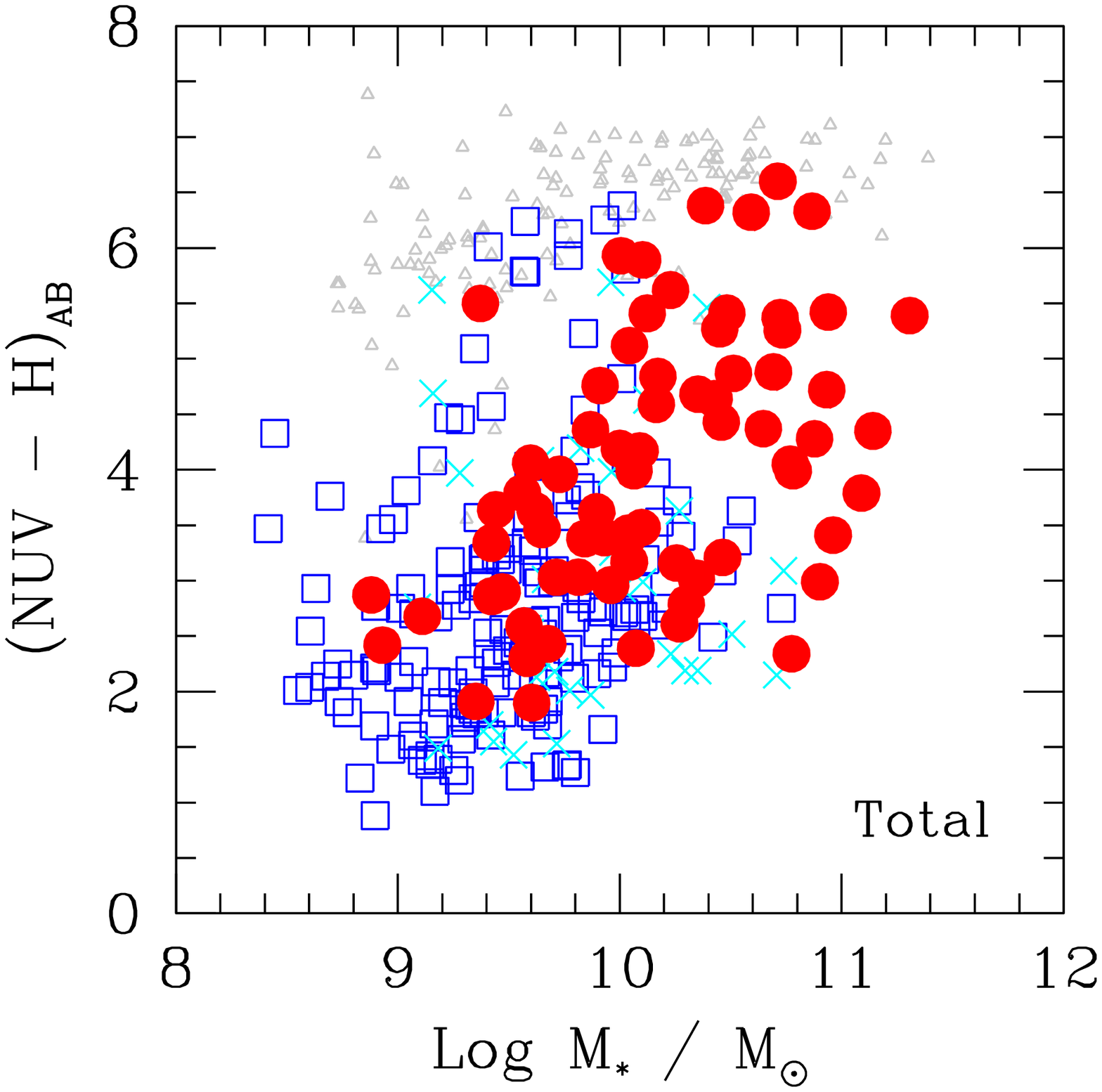}
\includegraphics[width=0.32\textwidth]{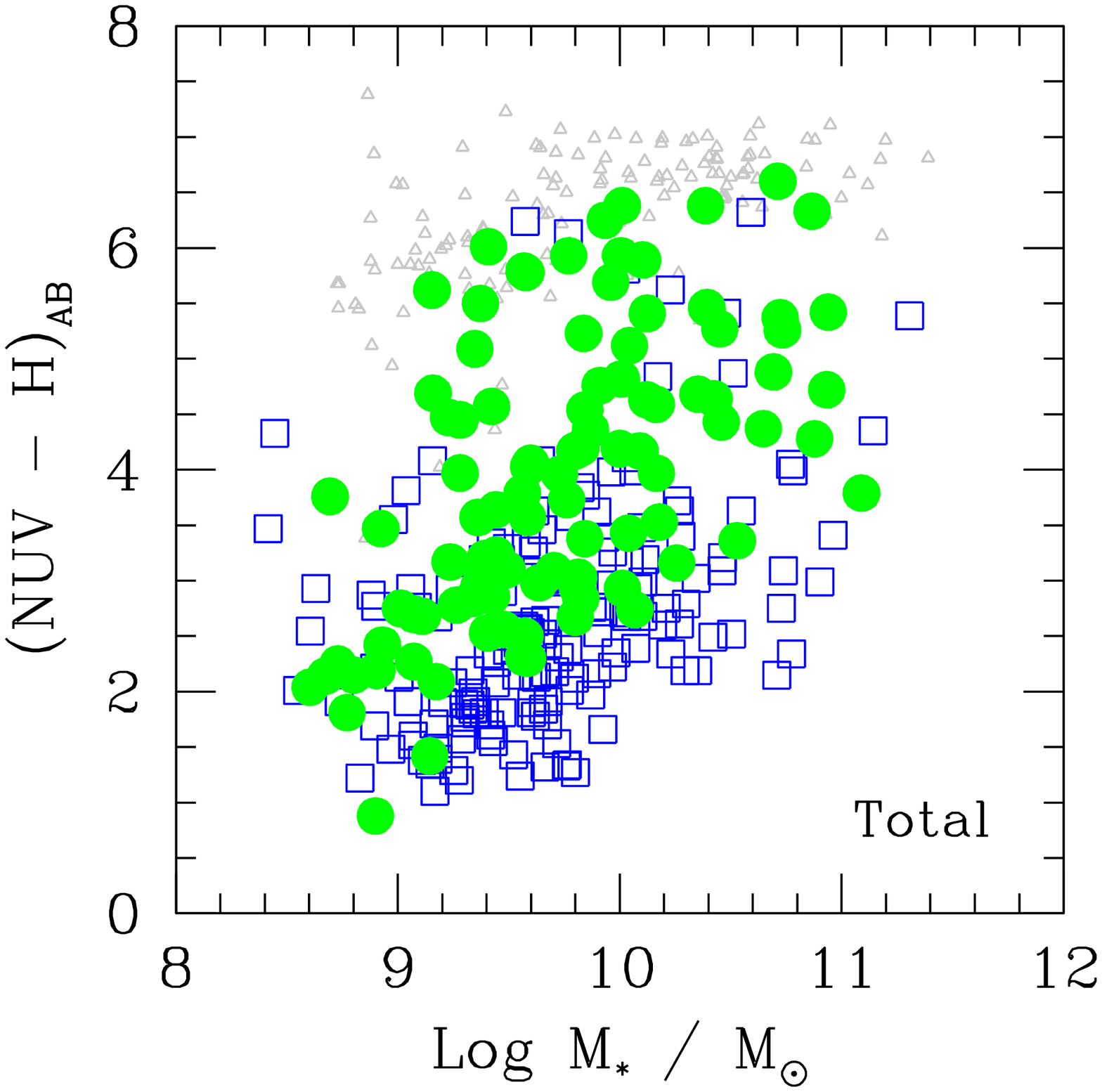}
\includegraphics[width=0.32\textwidth]{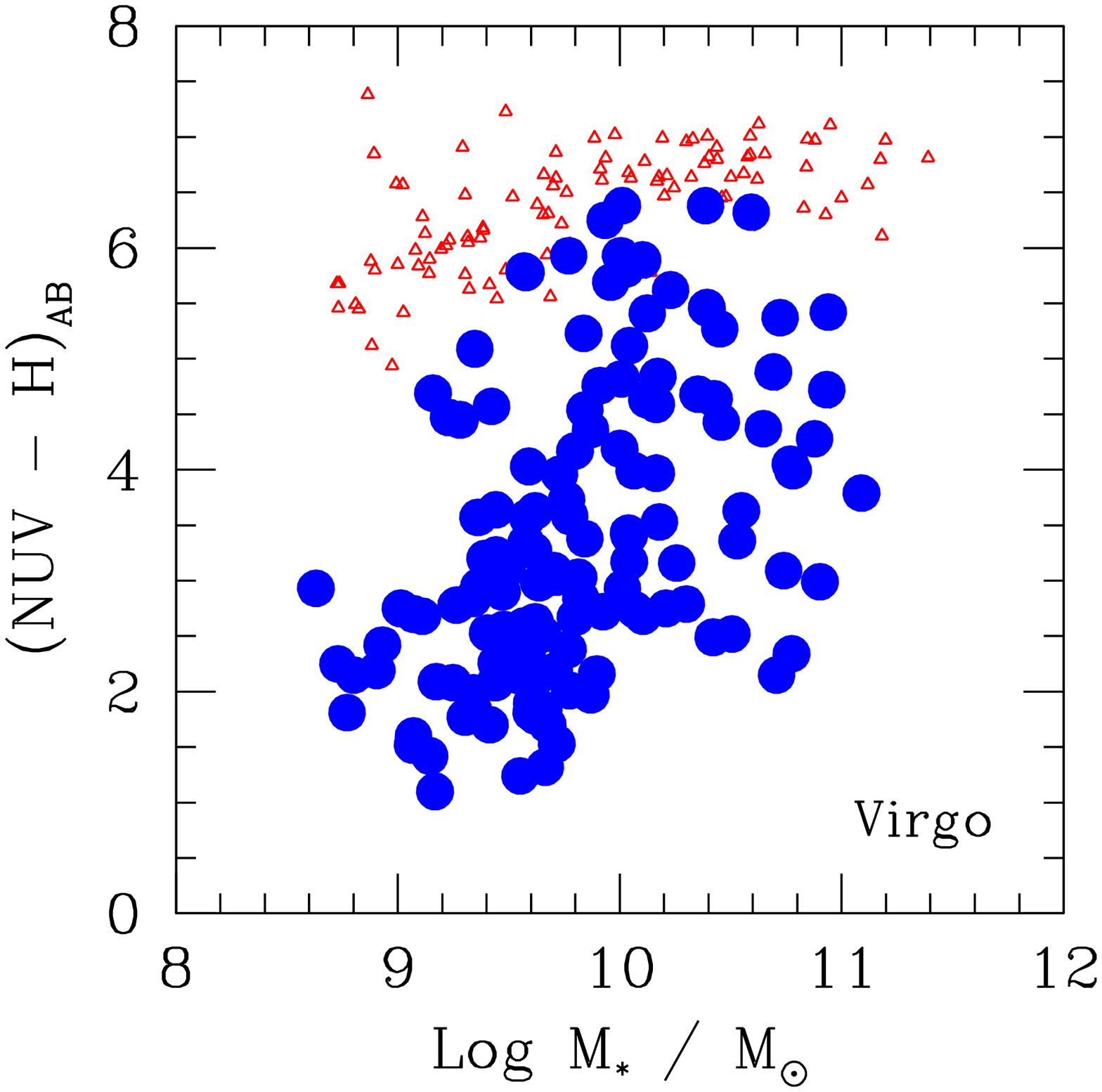}
\includegraphics[width=0.32\textwidth]{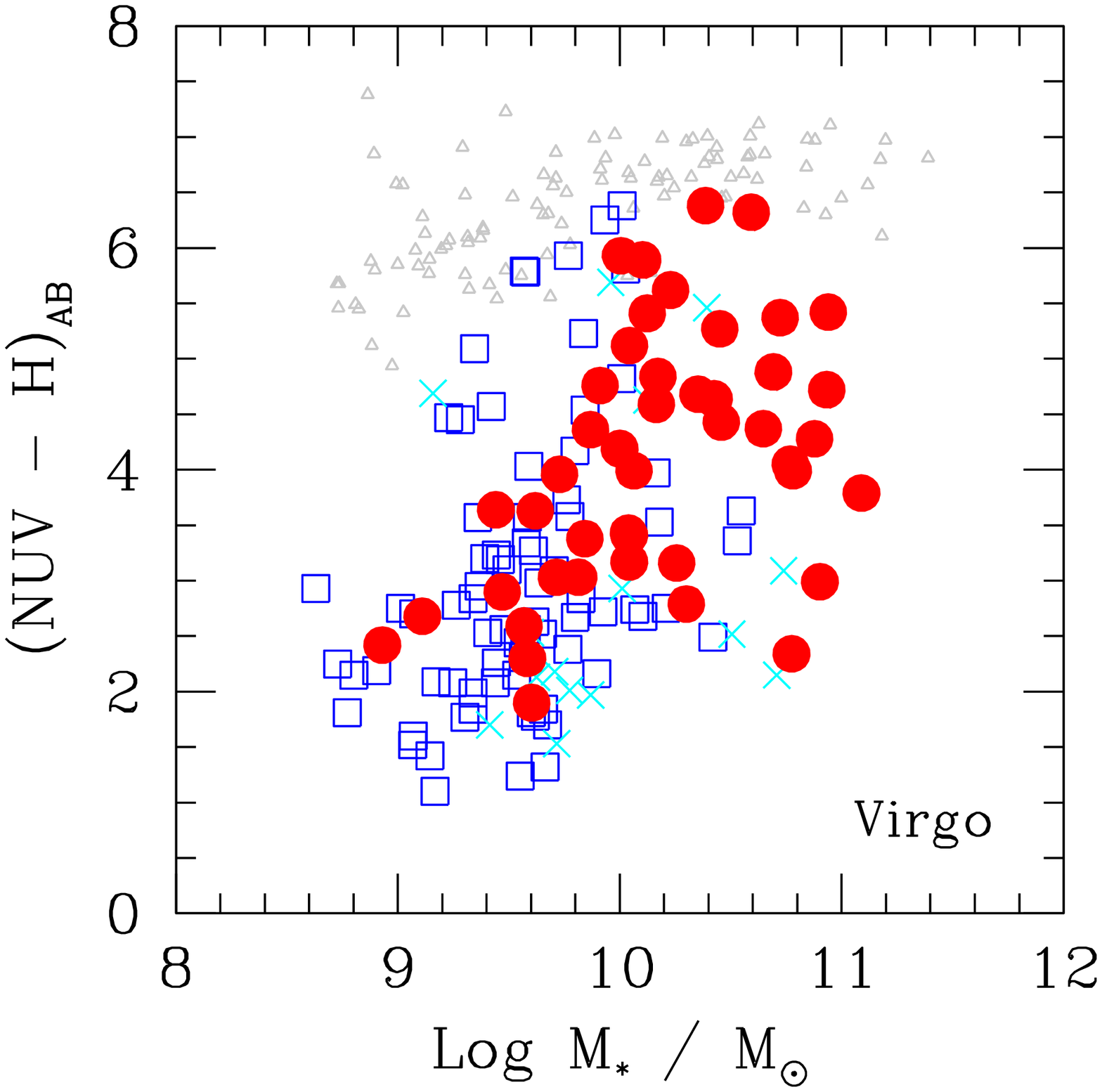}
\includegraphics[width=0.32\textwidth]{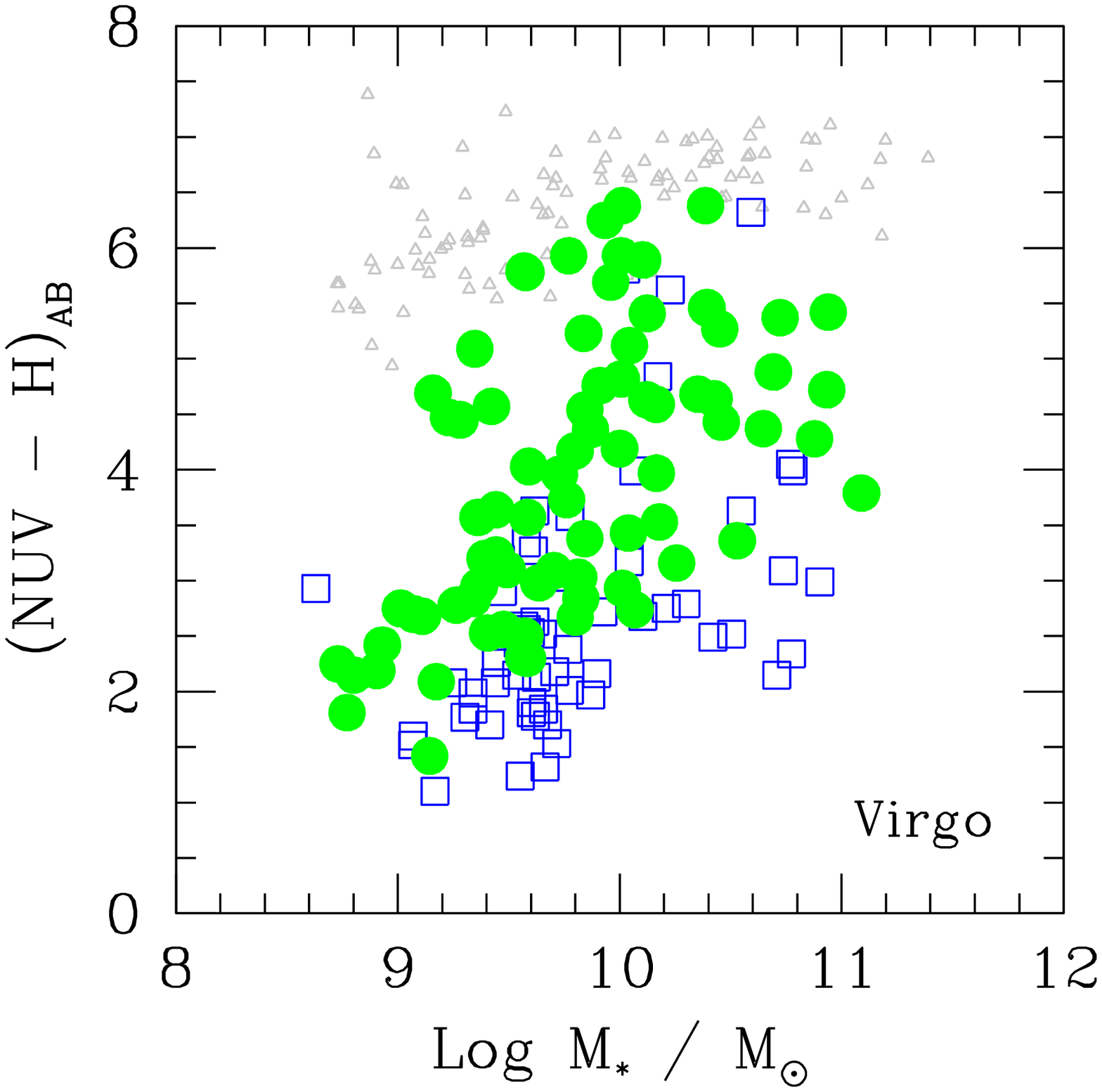}
\includegraphics[width=0.32\textwidth]{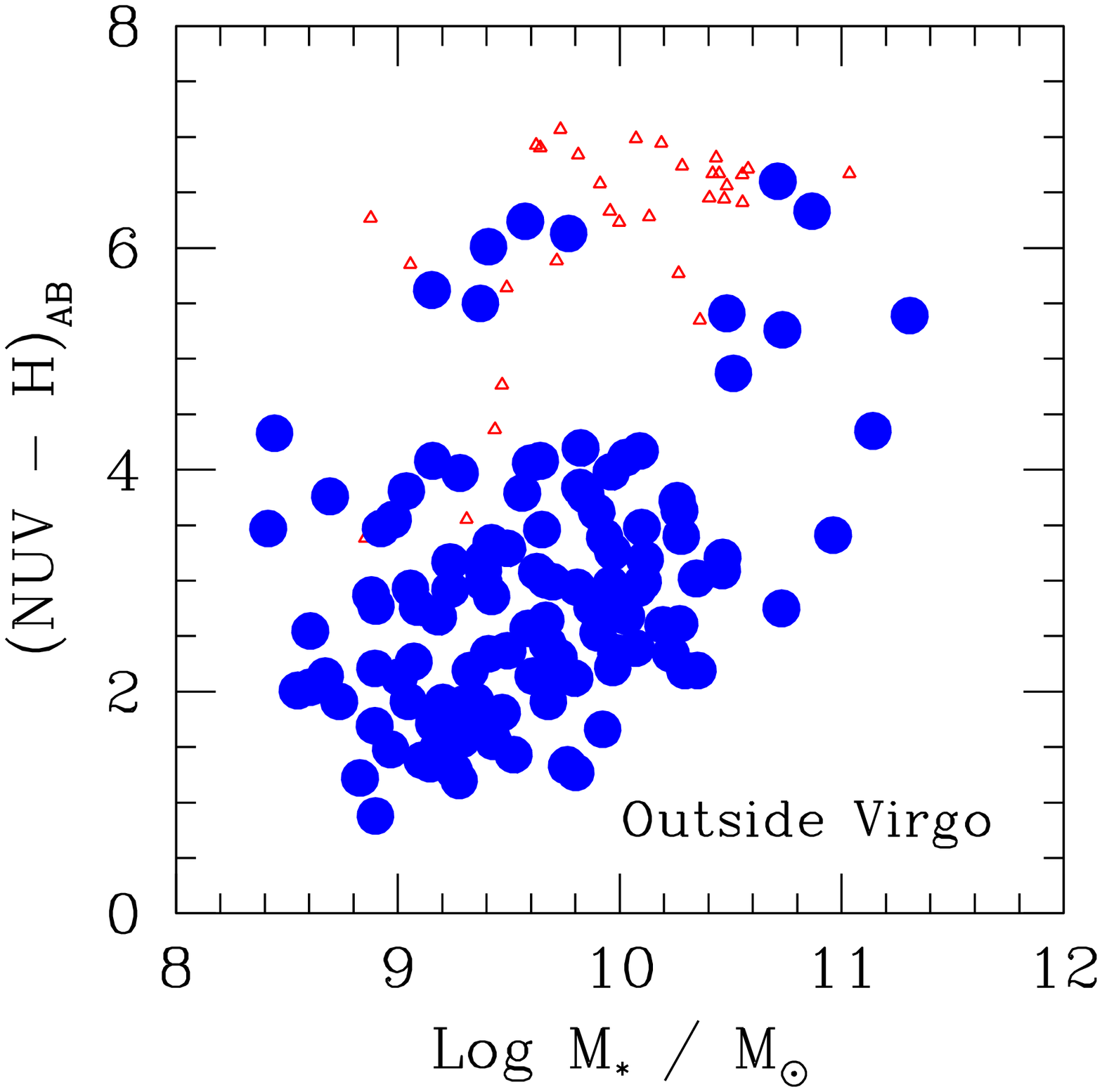}
\includegraphics[width=0.32\textwidth]{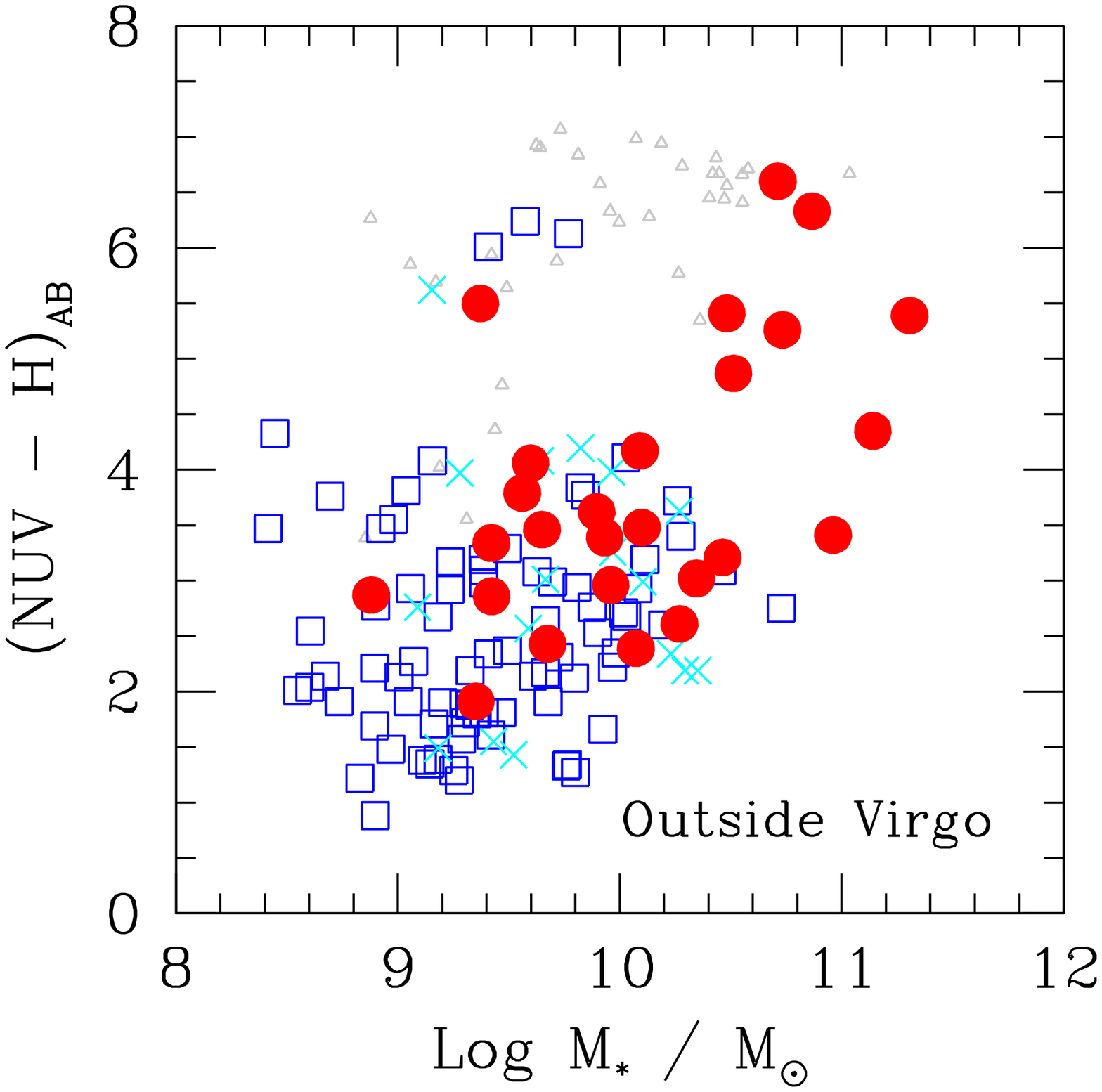}
\includegraphics[width=0.32\textwidth]{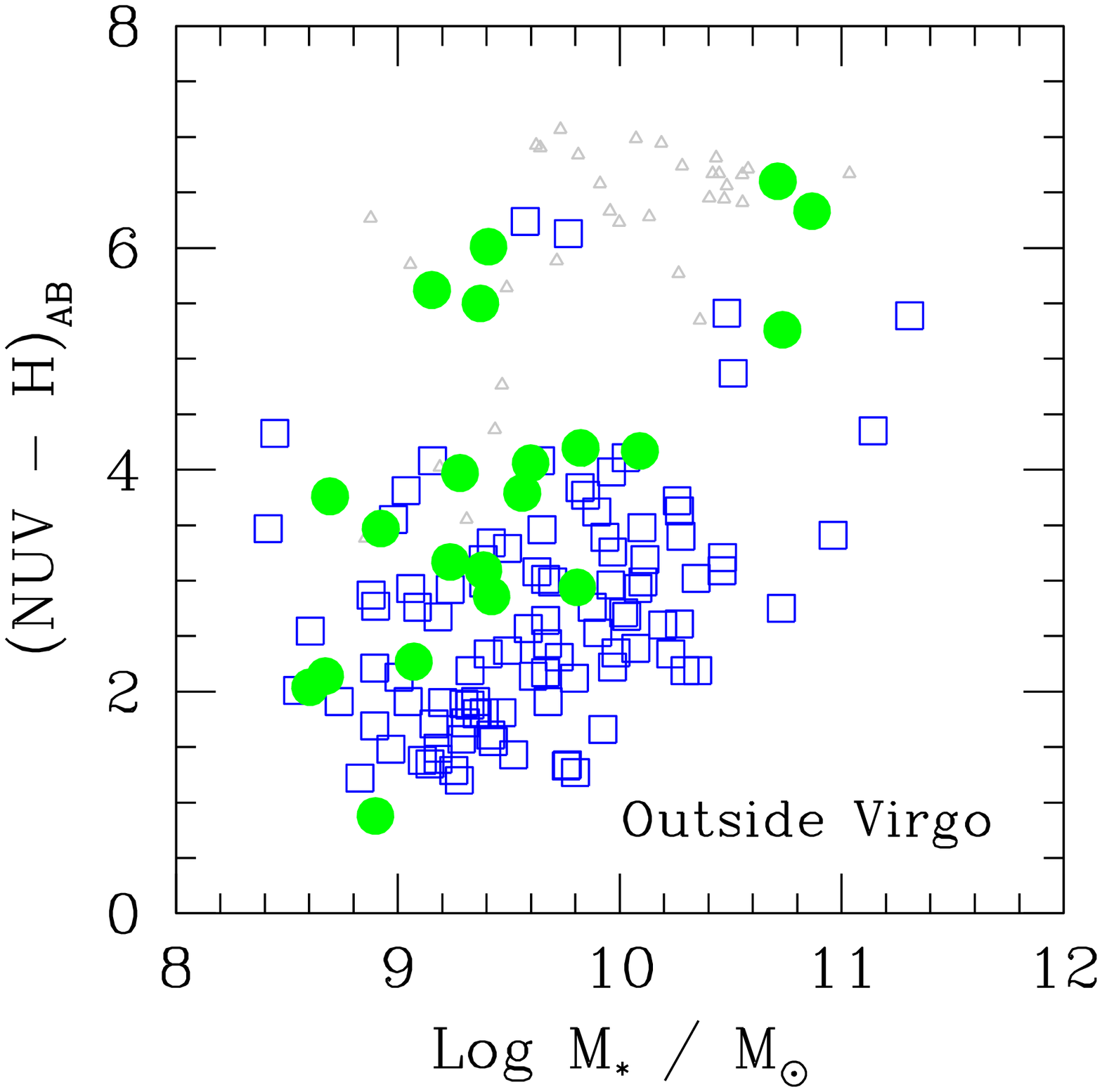}
\end{center}
\caption{
The \textit{NUV-H} vs. \textit{\mstar}\ colour-mass diagrams for galaxies comprising the total sample (\textit{top row}), the Virgo cluster (\textit{middle row}) and `field' environment (\textit{bottom row}). 
\textit{Left column}: Early-types (red triangles) form a red sequence and late-types (filled blue circles) form a blue cloud. 
\textit{Center column}: Objects displaying AGN-like behavior (filled red circles), \hii star-forming regions (blue open squares) or composites of the two (cyan crosses) are highlighted. 
\textit{Right column}: Galaxies are classified as \hi deficient (filled green circles) and non-\hi deficient (open blue squares). 
In the latter two columns, galaxies without nuclear classification or \hi observations are considered normal disks (blue open squares), and early-type galaxies (grey triangles) are retained only as a visual aid.}\label{fig:plots}
\end{figure*}

\section{Results}
The \textit{NUV-H} vs. \textit{\mstar}\ colour-mass diagrams for the total sample are presented in Fig.~\ref{fig:plots} (top row). 
The bimodality in the galaxy population is clearly evident, with early-type galaxies mainly segregated in the red sequence and late-types occupying the blue cloud.
As already shown by \cite{bergh2007} in optical and \cite{wyder2007} in UV, the two sequences are not well separated, since a  significant number of disk galaxies lie in the region between the two clouds (i.e., the transition region, $NUV-H >\ \sim$4.5 mag) and even in the red sequence.
This supports the common idea that galaxies start their journey towards the red sequence as disks.
Therefore, in the rest of this paper we focus our attention on the properties of late-type galaxies to try to understand the mechanisms behind the quenching of the star formation\footnote{We also note that a few early-type galaxies lie well outside the red sequence. 
These are misclassified or peculiar objects. Even if future investigation should reclassify them to be late-type galaxies, their exclusion does not affect the outcome of this work.}.   

To investigate the properties of galaxies in the transition region, we split our sample according to nuclear activity and \hi content (Fig.~\ref{fig:plots}, top row, central and right panel, respectively).
For $\mstar \gtrsim 10^{10}$ \Msolar, AGN-host spirals are uniformly distributed across the range of colour (consistent with \citealp{martin2007} and \citealp{schawinski2007}), whereas 
for lower stellar masses they are segregated in the blue sequence. 
In other words, {\it AGNs are not preferentially found in the transition region.} 
In addition, at fixed stellar mass, AGNs do not tend to be redder (i.e., have a lower specific star formation rate) than galaxies with \hii star-forming nuclei or composite systems. 
Thus,  we do not find any clear evidence of a suppression of the star formation in AGNs. 
On the contrary, quenched spirals are generally characterized by a low atomic hydrogen content. Unlike the AGN-hosts, which occupy the transition region only for masses $\mstar \gtrsim 10^{10}$ \Msolar, the \hi deficient galaxies tend to lie at the red edge of the blue sequence and in the transition region, independent of the stellar mass. More importantly, at fixed stellar mass, {\rm \hi}deficient systems have a typical colour at least 1 mag redder than gas-rich systems, suggesting that the gas depletion is really accompanied by a suppression of the star formation.
If this is the case, transition galaxies should be more 
frequent in high density environments, since the 
\hi deficiency is usually associated with a truncation of the gas disk via environmental effects \citep{cayatte1994}. In order to test this scenario, 
we split our sample depending upon whether the galaxies inhabit the Virgo cluster or not (see Fig.~\ref{fig:plots} middle and bottom rows).
A comparison of the cluster and `field' colour-mass diagrams clearly shows that very few disk galaxies have their star formation suppressed outside the cluster environment.
The vast majority of transition spirals is in fact composed of \hi deficient cluster galaxies.
In addition, almost 50\% of transition disks outside Virgo reside in groups or pairs, supporting the idea that the loss of gas and quenching of the star formation is related to the environment.
We note that some transition galaxies in our sample are not \hi deficient. These appear to be peculiar objects and their properties will be discussed in a future paper.

The significant difference found in the colour-mass distribution of galaxies in and outside Virgo is consistent with \cite{bergh2007}, who showed that objects in small groups differ from cluster galaxies and are more similar to isolated systems. 
However, we notice that in our `field' sample, the separation between the blue and the red sequence is more clear-cut than in optical colour-magnitude relations\footnote{For example, a K-S test shows that there is $<$0.01 \% probability that the $NUV-H$ colour distribution of 
our field sample has the same shape of the $U-B$ colour distribution in the \cite{bergh2007} `field+groups' sample.}.
The use of UV to near-infrared colours is in fact crucial to clearly show the environmental dependence of the transition population in the local universe.

Finally, our interpretation is additionally supported by the fact that (1) the fraction of AGN-host galaxies does not vary strongly with the environment (e.g., \citealp{miller2003,kauffmann2004}), and (2) also in the field, at fixed stellar mass, AGN-host spirals 
do not show a lower specific star formation rate than \hii galaxies. Both 
observational lines of evidence do in fact contrast with what is expected if quenching was via AGN feedback, favouring environment as the main mechanism responsible 
for the suppression of the star formation in our sample. 


\section{Discussion \& Conclusion}

In this Letter we have shown for the first time that, at all stellar masses, transition spirals tend to be \hi deficient galaxies, preferentially found in high-density environments. Although we confirm that the fraction of AGN-host spirals peaks in the transition region (see also \citealp{martin2007,schawinski2007}), we demonstrated that, at fixed stellar mass, AGNs 
are not associated with a systematic decrease in the star formation activity 
whereas \hi deficient galaxies are redder than gas-rich systems.
Thus, environmental effects could explain the suppression of the star formation over the whole range of stellar masses here investigated and no direct connection between AGN activity and 
quenching of the star formation is observed in our sample.

We note that biases in the classification of nuclear activity could be introduced by using optical emission lines.
On one hand, AGN activity in blue sequence galaxies could be totally obscured or outshone by star formation. 
This would imply an underestimate of the AGN fraction in the blue sequence, thus reinforcing our results.
On the other hand, \cite{sta2008} have recently shown that old stellar populations can provide enough photons to ionize the 
inter-stellar medium, mimicking AGN-type galaxies. This seems to be particularly important for LINER-like systems, which represent a large 
fraction ($\sim$56\%) of our AGN-host late-type galaxies. 
In this case, the observed peak of `so-called' nuclear activity in the transition region would merely be a consequence of the fact 
that transition spirals have an older stellar population than blue sequence disks, again supporting our results. 

The fact that the quenching of the star formation in transition galaxies is driven by the environment is not completely surprising, considering the plethora of observations supporting the significant impact of the environment on galaxy evolution (e.g., \citealp{dressler1980,boselli2006}). 
However, the use of \hi data to discriminate between the effects of AGN feedback and environment on star formation 
not only provides additional insights on the evolutionary history of spiral galaxies outside the blue sequence, but also shows that 
a physical link between AGN feedback and quenching may not be assumed from a correlation between nuclear activity and position in the colour-mass diagram.

Finally, we remind the reader that, since the AGN activity \citep{ueda2003}, the gas content of galaxies, and the properties of the large scale structure  evolve significantly with the age of the universe, the results obtained here cannot be blindly extended to high-redshift samples.
In fact, it would not be surprising if the dominant mechanism driving the quenching varies with redshift.
Although further work is required to shed additional light on the growth of the red sequence, we assert that environmental mechanisms 
have to be taken into account in the quest to better understand galaxy evolution in the local universe.

\section*{Acknowledgments}
We are greatly indebted to Alessandro Boselli for providing part of  
the data of the HRS before publication and
for useful discussions.
We wish to thank Barbara Catinella \& Jonathan Davies for useful  
comments and the referee, Sidney van den Bergh, for his comments and criticism which 
led us to improve the clarity of this paper.
We are supported by the UK Science and Technology Facilities Council.
This publication makes use of data from 2MASS, which is a joint project of the University of  
Massachusetts and the IPAC/Caltech, funded by the NASA and the NSF, and from 
the GALEX mission, developed in cooperation with the  
CNES-France and the Korean Ministry  
of Science and Technology.
This research has made use of the NED, which is operated by the JPL, Caltech, under  
contract to NASA and of the GOLDMine data base.
\bibliography{migration}

\begin{thebibliography}{}

\bibitem[\protect\citeauthoryear{{Baugh}}{{Baugh}}{2006}]{baugh2006}
{Baugh} C.~M.,  2006, Reports on Progress in Physics, 69, 3101

\bibitem[\protect\citeauthoryear{{Bell}, {McIntosh}, {Katz} \&
  {Weinberg}}{{Bell} et~al.}{2003}]{bell2003}
{Bell} E.~F.,  {McIntosh} D.~H.,  {Katz} N.,    {Weinberg} M.~D.,  2003, \apjs,
  149, 289

\bibitem[\protect\citeauthoryear{{Bell}, {Zheng}, {Papovich}, {Borch}, {Wolf}
  \& {Meisenheimer}}{{Bell} et~al.}{2007}]{bell2007}
{Bell} E.~F.,  {Zheng} X.~Z.,  {Papovich} C.,  {Borch} A.,  {Wolf} C.,
  {Meisenheimer} K.,  2007, \apj, 663, 834

\bibitem[\protect\citeauthoryear{{Binggeli}, {Sandage} \& {Tammann}}{{Binggeli}
  et~al.}{1985}]{binggeli1985}
{Binggeli} B.,  {Sandage} A.,    {Tammann} G.~A.,  1985, \aj, 90, 1681

\bibitem[\protect\citeauthoryear{{Boselli}, {Boissier}, {Cortese} \&
  {Gavazzi}}{{Boselli} et~al.}{2008}]{boselli2008}
{Boselli} A.,  {Boissier} S.,  {Cortese} L.,    {Gavazzi} G.,  2008, \apj, 674,
  742

\bibitem[\protect\citeauthoryear{{Boselli}, {Eales}, {Bendo}, {Buat},
  {Chanial}, {Cortese}, {Davies}, {Baes} \& {et al.}}{{Boselli}
  et~al.}{2009}]{boselli2009}
{Boselli} A.,  {Eales} S.,  {Bendo} G.,  {Buat} V.,  {Chanial} P.,  {Cortese}
  L.,  {Davies} J.,  {Baes} M.,    {et al.} 2009, \pasp,\ submitted

\bibitem[\protect\citeauthoryear{{Boselli} \& {Gavazzi}}{{Boselli} \&
  {Gavazzi}}{2006}]{boselli2006}
{Boselli} A.,  {Gavazzi} G.,  2006, \pasp, 118, 517

\bibitem[\protect\citeauthoryear{{Cayatte}, {Kotanyi}, {Balkowski} \& {van
  Gorkom}}{{Cayatte} et~al.}{1994}]{cayatte1994}
{Cayatte} V.,  {Kotanyi} C.,  {Balkowski} C.,    {van Gorkom} J.~H.,  1994,
  \aj, 107, 1003

\bibitem[\protect\citeauthoryear{{Cortese}, {Boselli}, {Buat}, {Gavazzi},
  {Boissier}, {Gil de Paz}, {Seibert}, {Madore} \& {Martin}}{{Cortese}
  et~al.}{2006}]{cortese2006a}
{Cortese} L.,  {Boselli} A.,  {Buat} V.,  {Gavazzi} G.,  {Boissier} S.,  {Gil
  de Paz} A.,  {Seibert} M.,  {Madore} B.~F.,    {Martin} D.~C.,  2006, \apj,
  637, 242

\bibitem[\protect\citeauthoryear{{Cortese}, {Boselli}, {Franzetti}, {Decarli},
  {Gavazzi}, {Boissier} \& {Buat}}{{Cortese} et~al.}{2008}]{cortese2008}
{Cortese} L.,  {Boselli} A.,  {Franzetti} P.,  {Decarli} R.,  {Gavazzi} G.,
  {Boissier} S.,    {Buat} V.,  2008, \mnras, 386, 1157

\bibitem[\protect\citeauthoryear{{Cortese}, {Gavazzi}, {Boselli}, {Franzetti},
  {Kennicutt}, {O'Neil} \& {Sakai}}{{Cortese} et~al.}{2006}]{cortese2006b}
{Cortese} L.,  {Gavazzi} G.,  {Boselli} A.,  {Franzetti} P.,  {Kennicutt}
  R.~C.,  {O'Neil} K.,    {Sakai} S.,  2006, \aap, 453, 847

\bibitem[\protect\citeauthoryear{{Croton}, {Springel}, {White}, {De Lucia},
  {Frenk}, {Gao}, {Jenkins}, {Kauffmann}, {Navarro} \& {Yoshida}}{{Croton}
  et~al.}{2006}]{croton2006}
{Croton} D.~J.,  {Springel} V.,  {White} S.~D.~M.,  {De Lucia} G.,  {Frenk}
  C.~S.,  {Gao} L.,  {Jenkins} A.,  {Kauffmann} G.,  {Navarro} J.~F.,
  {Yoshida} N.,  2006, \mnras, 365, 11

\bibitem[\protect\citeauthoryear{{de Vaucouleurs}, {de Vaucouleurs}, {Corwin}
  Jr., {Buta}, {Paturel} \& {Fouque}}{{de Vaucouleurs} et~al.}{1991}]{vauc1991}
{de Vaucouleurs} G.,  {de Vaucouleurs} A.,  {Corwin} Jr. H.~G.,  {Buta} R.~J.,
  {Paturel} G.,    {Fouque} P.,  1991, {Third Reference Catalogue of Bright
  Galaxies}.
Volume 1-3, XII, 2069 pp.~7 ~Springer-Verlag Berlin Heidelberg New York

\bibitem[\protect\citeauthoryear{{Decarli}, {Gavazzi}, {Arosio}, {Cortese},
  {Boselli}, {Bonfanti} \& {Colpi}}{{Decarli} et~al.}{2007}]{decarli2007}
{Decarli} R.,  {Gavazzi} G.,  {Arosio} I.,  {Cortese} L.,  {Boselli} A.,
  {Bonfanti} C.,    {Colpi} M.,  2007, \mnras, 381, 136

\bibitem[\protect\citeauthoryear{{Dressler}}{{Dressler}}{1980}]{dressler1980}
{Dressler} A.,  1980, \apj, 236, 351

\bibitem[\protect\citeauthoryear{{Faber}, {Willmer}, {Wolf}, {Koo}, {Weiner},
  {Newman}, {Im}, {Coil} \& {Conroy}}{{Faber} et~al.}{2007}]{faber2007}
{Faber} S.~M.,  {Willmer} C.~N.~A.,  {Wolf} C.,  {Koo} D.~C.,  {Weiner} B.~J.,
  {Newman} J.~A.,  {Im} M.,  {Coil} A.~L.,    {Conroy} C.,  2007, \apj, 665,
  265

\bibitem[\protect\citeauthoryear{{Forman}, {Jones}, {Churazov}, {Markevitch},
  {Nulsen}, {Vikhlinin}, {Begelman}, {B{\"o}hringer}, {Eilek}, {Heinz},
  {Kraft}, {Owen} \& {Pahre}}{{Forman} et~al.}{2007}]{forman2007}
{Forman} W.,  {Jones} C.,  {Churazov} E.,  {Markevitch} M.,  {Nulsen} P.,
  {Vikhlinin} A.,  {Begelman} M.,  {B{\"o}hringer} H.,  {Eilek} J.,  {Heinz}
  S.,  {Kraft} R.,  {Owen} F.,    {Pahre} M.,  2007, \apj, 665, 1057

\bibitem[\protect\citeauthoryear{{Fujita}}{{Fujita}}{2004}]{fujita2004}
{Fujita} Y.,  2004, \pasj, 56, 29

\bibitem[\protect\citeauthoryear{{Gavazzi}, {Boselli}, {Donati}, {Franzetti} \&
  {Scodeggio}}{{Gavazzi} et~al.}{2003}]{gav2003}
{Gavazzi} G.,  {Boselli} A.,  {Donati} A.,  {Franzetti} P.,    {Scodeggio} M.,
  2003, \aap, 400, 451

\bibitem[\protect\citeauthoryear{{Gavazzi}, {Boselli}, {Scodeggio}, {Pierini}
  \& {Belsole}}{{Gavazzi} et~al.}{1999}]{gav1999}
{Gavazzi} G.,  {Boselli} A.,  {Scodeggio} M.,  {Pierini} D.,    {Belsole} E.,
  1999, \mnras, 304, 595

\bibitem[\protect\citeauthoryear{{Georgakakis}, {Nandra}, {Yan}, {Willner},
  {Lotz}, {Pierce}, {Cooper}, {Laird}, {Koo}, {Barmby}, {Newman}, {Primack} \&
  {Coil}}{{Georgakakis} et~al.}{2008}]{geo2008}
{Georgakakis} A.,  {Nandra} K.,  {Yan} R.,  {Willner} S.~P.,  {Lotz} J.~M.,
  {Pierce} C.~M.,  {Cooper} M.~C.,  {Laird} E.~S.,  {Koo} D.~C.,  {Barmby} P.,
  {Newman} J.~A.,  {Primack} J.~R.,    {Coil} A.~L.,  2008, \mnras, 385, 2049

\bibitem[\protect\citeauthoryear{{Gunn} \& {Gott}}{{Gunn} \&
  {Gott}}{1972}]{gunngott1972}
{Gunn} J.~E.,  {Gott} J.~R.~I.,  1972, \apj, 176, 1

\bibitem[\protect\citeauthoryear{{Haynes} \& {Giovanelli}}{{Haynes} \&
  {Giovanelli}}{1984}]{haynes1984}
{Haynes} M.~P.,  {Giovanelli} R.,  1984, \aj, 89, 758

\bibitem[\protect\citeauthoryear{{Jarrett}, {Chester}, {Cutri}, {Schneider} \&
  {Huchra}}{{Jarrett} et~al.}{2003}]{jarrett2003}
{Jarrett} T.~H.,  {Chester} T.,  {Cutri} R.,  {Schneider} S.~E.,    {Huchra}
  J.~P.,  2003, \aj, 125, 525

\bibitem[\protect\citeauthoryear{{Kauffmann}, {White}, {Heckman}, {M{\'e}nard},
  {Brinchmann}, {Charlot}, {Tremonti} \& {Brinkmann}}{{Kauffmann}
  et~al.}{2004}]{kauffmann2004}
{Kauffmann} G.,  {White} S.~D.~M.,  {Heckman} T.~M.,  {M{\'e}nard} B.,
  {Brinchmann} J.,  {Charlot} S.,  {Tremonti} C.,    {Brinkmann} J.,  2004,
  \mnras, 353, 713

\bibitem[\protect\citeauthoryear{{Kennicutt}
  Jr.}{{Kennicutt}}{1983}]{kennicutt1983}
{Kennicutt} Jr. R.~C.,  1983, \aj, 88, 483

\bibitem[\protect\citeauthoryear{{Kroupa}, {Tout} \& {Gilmore}}{{Kroupa}
  et~al.}{1993}]{kroupa1993}
{Kroupa} P.,  {Tout} C.~A.,    {Gilmore} G.,  1993, \mnras, 262, 545

\bibitem[\protect\citeauthoryear{{Martin}, {Fanson}, {Schiminovich},
  {Morrissey}, {Friedman}, {Barlow}, {Conrow}, {Grange} \& {et al.}}{{Martin}
  et~al.}{2005}]{martin2005}
{Martin} D.~C.,  {Fanson} J.,  {Schiminovich} D.,  {Morrissey} P.,  {Friedman}
  P.~G.,  {Barlow} T.~A.,  {Conrow} T.,  {Grange} R.,    {et al.} 2005, \apjl,
  619, L1

\bibitem[\protect\citeauthoryear{{Martin}, {Wyder}, {Schiminovich}, {Barlow},
  {Forster}, {Friedman}, {Morrissey}, {Neff} \& {et al.}}{{Martin}
  et~al.}{2007}]{martin2007}
{Martin} D.~C.,  {Wyder} T.~K.,  {Schiminovich} D.,  {Barlow} T.~A.,  {Forster}
  K.,  {Friedman} P.~G.,  {Morrissey} P.,  {Neff} S.~G.,    {et al.} 2007,
  \apjs, 173, 342

\bibitem[\protect\citeauthoryear{{Merritt}}{{Merritt}}{1984}]{merritt1984}
{Merritt} D.,  1984, \apj, 276, 26

\bibitem[\protect\citeauthoryear{{Miller}, {Nichol}, {G{\'o}mez}, {Hopkins} \&
  {Bernardi}}{{Miller} et~al.}{2003}]{miller2003}
{Miller} C.~J.,  {Nichol} R.~C.,  {G{\'o}mez} P.~L.,  {Hopkins} A.~M.,
  {Bernardi} M.,  2003, \apj, 597, 142

\bibitem[\protect\citeauthoryear{{Moore}, {Katz}, {Lake}, {Dressler} \&
  {Oemler}}{{Moore} et~al.}{1996}]{moore1996}
{Moore} B.,  {Katz} N.,  {Lake} G.,  {Dressler} A.,    {Oemler} A.,  1996,
  \nat, 379, 613

\bibitem[\protect\citeauthoryear{{Okamoto}, {Nemmen} \& {Bower}}{{Okamoto}
  et~al.}{2008}]{okamoto2008}
{Okamoto} T.,  {Nemmen} R.~S.,    {Bower} R.~G.,  2008, \mnras, 385, 161

\bibitem[\protect\citeauthoryear{{Schawinski}, {Thomas}, {Sarzi}, {Maraston},
  {Kaviraj}, {Joo}, {Yi} \& {Silk}}{{Schawinski} et~al.}{2007}]{schawinski2007}
{Schawinski} K.,  {Thomas} D.,  {Sarzi} M.,  {Maraston} C.,  {Kaviraj} S.,
  {Joo} S.-J.,  {Yi} S.~K.,    {Silk} J.,  2007, \mnras, 382, 1415

\bibitem[\protect\citeauthoryear{{Schlegel}, {Finkbeiner} \&
  {Davis}}{{Schlegel} et~al.}{1998}]{schlegel1998}
{Schlegel} D.~J.,  {Finkbeiner} D.~P.,    {Davis} M.,  1998, \apj, 500, 525

\bibitem[\protect\citeauthoryear{{Solanes}, {Giovanelli} \& {Haynes}}{{Solanes}
  et~al.}{1996}]{solanes1996}
{Solanes} J.~M.,  {Giovanelli} R.,    {Haynes} M.~P.,  1996, \apj, 461, 609

\bibitem[\protect\citeauthoryear{{Springob}, {Haynes}, {Giovanelli} \&
  {Kent}}{{Springob} et~al.}{2005}]{springob2005}
{Springob} C.~M.,  {Haynes} M.~P.,  {Giovanelli} R.,    {Kent} B.~R.,  2005,
  \apjs, 160, 149

\bibitem[\protect\citeauthoryear{{Stasi{\'n}ska}, {Asari}, {Fernandes},
  {Gomes}, {Schlickmann}, {Mateus}, {Schoenell} \& {Sodr{\'e}}
  Jr.}{{Stasi{\'n}ska} et~al.}{2008}]{sta2008}
{Stasi{\'n}ska} G.,  {Asari} N.~V.,  {Fernandes} R.~C.,  {Gomes} J.~M.,
  {Schlickmann} M.,  {Mateus} A.,  {Schoenell} W.,    {Sodr{\'e}} Jr. L.,
  2008, \mnras, 391, L29

\bibitem[\protect\citeauthoryear{{Tully}, {Mould} \& {Aaronson}}{{Tully}
  et~al.}{1982}]{tully1982}
{Tully} R.~B.,  {Mould} J.~R.,    {Aaronson} M.,  1982, \apj, 257, 527

\bibitem[\protect\citeauthoryear{{Ueda}, {Akiyama}, {Ohta} \& {Miyaji}}{{Ueda}
  et~al.}{2003}]{ueda2003}
{Ueda} Y.,  {Akiyama} M.,  {Ohta} K.,    {Miyaji} T.,  2003, \apj, 598, 886

\bibitem[\protect\citeauthoryear{{van den Bergh}}{{van den
  Bergh}}{2007}]{bergh2007}
{van den Bergh} S.,  2007, \aj, 134, 1508

\bibitem[\protect\citeauthoryear{{Whitmore}, {Gilmore} \& {Jones}}{{Whitmore}
  et~al.}{1993}]{whitmore1993}
{Whitmore} B.~C.,  {Gilmore} D.~M.,    {Jones} C.,  1993, \apj, 407, 489

\bibitem[\protect\citeauthoryear{{Wyder}, {Martin}, {Schiminovich}, {Seibert},
  {Budav{\'a}ri}, {Treyer}, {Barlow}, {Forster} \& {et al.}}{{Wyder}
  et~al.}{2007}]{wyder2007}
{Wyder} T.~K.,  {Martin} D.~C.,  {Schiminovich} D.,  {Seibert} M.,
  {Budav{\'a}ri} T.,  {Treyer} M.~A.,  {Barlow} T.~A.,  {Forster} K.,    {et
  al.} 2007, \apjs, 173, 293

\bibitem[\protect\citeauthoryear{{Xu} \& {Buat}}{{Xu} \& {Buat}}{1995}]{xu1995}
{Xu} C.,  {Buat} V.,  1995, \aap, 293, L65

\bibitem[\protect\citeauthoryear{{York}, {Adelman}, {Anderson} Jr., {Anderson},
  {Annis}, {Bahcall}, {Bakken}, {Barkhouser} \& {et al.}}{{York}
  et~al.}{2000}]{sdss2000}
{York} D.~G.,  {Adelman} J.,  {Anderson} Jr. J.~E.,  {Anderson} S.~F.,  {Annis}
  J.,  {Bahcall} N.~A.,  {Bakken} J.~A.,  {Barkhouser} R.,    {et al.} 2000,
  \aj, 120, 1579

\end{thebibliography}
\end{document}